\documentclass[journal,twocolumn]{IEEEtran}
\usepackage[T1]{fontenc}
\usepackage{cite}
\usepackage{amsthm,amssymb,bm}
\usepackage[tbtags]{amsmath}
\usepackage{mathrsfs}
\usepackage{graphicx}
\usepackage{enumitem}
\usepackage{algorithm}  
\usepackage{algorithmic}  
\usepackage{array}
\usepackage{psfrag}
\usepackage{booktabs}
\usepackage{tabularx} 
\usepackage{xcolor}
\usepackage[caption=false,font=normalsize,labelfont=sf,textfont=sf]{subfig}
\usepackage{amsmath}
\usepackage{cuted}
\allowdisplaybreaks[4]
\UseRawInputEncoding
\ifCLASSINFOpdf
\else
\fi

\newcounter{MYtempeqncnt}

\begin{document}

\renewcommand{\qedsymbol}{}
%

\title{Low Time Complexity Near-Field  Channel and Position Estimations  }

\author{
	Xiyuan~Liu,~\IEEEmembership{Graduate Student Member,~IEEE,}
	Qingqing~Wu,~\IEEEmembership{Senior Member,~IEEE,}
	Rui~Wang,~\IEEEmembership{Senior Member,~IEEE,}
	Jun~Wu,~\IEEEmembership{Senior Member,~IEEE,}
	\thanks{X. Liu is with the College of Electronics and Information Engineering, Tongji University, Shanghai 201804, China (e-mail: 1910670@tongji.edu.cn).
		Q. Wu is with the Department of Electronic Engineering, Shanghai Jiao Tong University, 200240, China (e-mail: qingqingwu@sjtu.edu.cn).
		R. Wang is with the College of Electronics and Information Engineering, Tongji University, Shanghai 201804, China, and also with the Shanghai Institute of Intelligent Science and Technology, Tongji University, Shanghai 201804, China (e-mail: ruiwang@tongji.edu.cn).
		J. Wu is with the School of Computer Science, Fudan University, Shanghai 200433, China, and also with the Shanghai Qi Zhi Institute, Shanghai 200030, China (e-mail: wujun@fudan.edu.cn).}	
}

\maketitle

%



\IEEEtitleabstractindextext{%
	\begin{abstract}

		With the application of high-frequency communication and extremely large MIMO (XL-MIMO), the near-field effect has become increasingly apparent. The near-field channel estimation and position estimation problems both rely on the Angle of Arrival (AoA) and the Curvature of Arrival (CoA) estimation. However, in the near-field channel model, the coupling of AoA and CoA information poses a challenge to the estimation of the near-field channel. This paper proposes a Joint Autocorrelation and Cross-correlation (JAC) scheme to decouple AoA and CoA estimation. Based on the JAC scheme, we propose two specific near-field estimation algorithms, namely Inverse Sinc Function (JAC-ISF) and Gradient Descent (JAC-GD) algorithms. Finally, we analyzed the time complexity of the JAC scheme and the cramer-rao lower bound  (CRLB) for near-field position estimation. The simulation experiment results show that the algorithm designed based on JAC scheme can solve the problem of coupled CoA and AoA information in near-field estimation, thereby improving the algorithm performance. The JAC-GD algorithm shows significant performance in channel estimation and position estimation at different SNRs, snapshot points, and communication distances compared to other algorithms. This indicates that the JAC-GD algorithm can achieve more accurate channel and position estimation results while saving time overhead. 
		
	\end{abstract}
	
	}


\IEEEdisplaynontitleabstractindextext

%
\IEEEpeerreviewmaketitle

\section{Introduction}
Since the concept of MIMO was introduced, the scale of MIMO systems has been expanding, evolving from massive MIMO \cite{DBLP:journals/bell/Marzetta15}  to extremely large MIMO (XL-MIMO) \cite{9184098, DBLP:journals/dsp/BjornsonSWHM19, 8644126, 8948302, DBLP:journals/icl/WeiD22, DBLP:journals/tcom/0013Z0LHY24}. This progression is primarily driven by two key factors.  First, the frequency band of communication has expanded from the original sub-6 GHz to high-frequency bands like millimeter wave and terahertz \cite{DBLP:journals/comsur/WangYGZLZWHCHTLRTZSPH23, 9324910,9933498, 9665432}. Due to the limited scattering and diffraction capabilities of high-frequency signals \cite{9184098, 10149471}, XL-MIMO becomes crucial as it provides beamforming gain to compensate for path loss in high-frequency signal propagation. Simultaneously, the shorter wavelength of high-frequency signals allows the integration of larger MIMOs within a limited physical size. Second, the adoption of hybrid beamforming and intelligent reflecting surface technologies has significantly reduced the average cost of MIMO \cite{DBLP:journals/twc/ZhengZ23, DBLP:journals/twc/WuZ19}. Consequently, XL-MIMO is deemed necessary, feasible, and practical in the 6G communication scenario.

However, XL-MIMO can introduce significant near-field effects \cite{DBLP:conf/iccchina/PengZJLL23, DBLP:conf/acssc/TorresSB20, 9743350, 9903389}. It is widely acknowledged that the boundary between the far-field and near-field is defined by the Rayleigh distance, expressed as $\frac{2D^2}{\lambda}$ \cite{7942128}, where $D$ represents the array size, and $\lambda$ is the wavelength of the signal. In XL-MIMO and high-frequency communication scenarios, the larger $D$ and smaller $\lambda$ result in a substantially increased Rayleigh distance compared to traditional communication scenarios \cite{DBLP:conf/wcsp/WangZZTY21,DBLP:conf/pimrc/HuIW22}. This leads to more users being in the near-field range. The near-field effect introduces significant challenges to beamforming design, codebook design \cite{DBLP:journals/tcom/0013Z0LHY24, 2023Extremely}, and the beam training process \cite{10005200}, emerging as the primary bottleneck in current high-frequency communication and XL-MIMO scenarios.

The main distinction between the near-field and far-field lies in the fact that electromagnetic waves exhibit spherical wave characteristics in the near-field, as opposed to plane waves in the far-field \cite{DBLP:journals/icl/WeiD22}. Consequently, in the far-field, electromagnetic wave characteristics can be solely described through the angles of arrival (AoA) or departure (AoD).  In the near-field region, owing to the characteristics of spherical waves, the AoAs observed by antennas at different positions of the array vary \cite{7942128} \cite{10146329}. To elucidate the AoA variations, the introduction of the concept of curvature of arrival (CoA) becomes necessary to assist in delineating the spatial characteristics of electromagnetic waves in near-field channels \cite{DBLP:journals/tsp/Friedlander19a}. Generally, the CoA of spherical waves in the near-field region can be considered constant \cite{9184098}. Therefore, combining CoA information with AoA information at the reference antenna enables the description of the characteristics of the near-field channel.

However, the introduction of CoA significantly complicates the near-field problem compared to the far-field problem. There are two primary challenges in beam training within near-field scenarios. First, the joint estimation of CoA and AoA introduces a multiplicative time complexity, resulting in a quadratic form of the number of antennas. This complexity arises for the reason that the array's resolution for both AoA and CoA is  proportional to the array's  aperture, denoted as $\frac{D}{\lambda}$. In the context of XL-MIMO, the time complexity of separately estimating AoA or CoA is $\mathcal{O}(N)$, where $N$ represents the number of antennas in the array. When estimating AoA and CoA simultaneously, the time complexity becomes $\mathcal{O}(N^2)$ due to their combination in forming electromagnetic waves of different shapes. Given that XL-MIMO involves a larger value of $N$ compared to Massive MIMO, the near-field beam training problem becomes exceptionally time-consuming, exacerbated by the quadratic time complexity and larger $N$. Consequently, designing an efficient beam training scheme in the near-field becomes a considerable challenge. The second primary challenge faced by near-field beam training is the coupling of CoA and AoA information, presenting difficulties in orthogonal decomposition of the near-field space \cite{10476971}. 

However, if orthogonal decomposition cannot be achieved in the near field, it can lead to the following two main problems. The first problem is the further increase in the time complexity of near-field beamforming. Without the ability to orthogonally decompose space, it becomes impossible to explore the entire space with a finite number of beams without duplication or leakage. Consequently, it becomes necessary to repeat a certain area to enhance the spatial coverage of the beam training process, introducing significant redundancy and consuming additional time \cite{10132547}.  The second issue is that when estimating coupled variables, the two variables can affect each other, making the estimation problem more complex. This has had a serious impact on issues such as near-field codebook design, beam training, channel estimation, and position estimation.

To address these gaps, We need to search for received signal statistics that are only related to AoA or CoA. We search for such statistics based on the dual properties of time-domain signal processing and spatial signal processing to decouple near-field channel parameters. Based on these statistics, we design a scheme for near-field channel estimation and position estimation, and then based on this scheme, we can design specific estimation algorithms. Our main contributions are as follows:
\begin{itemize}
	\item 
	We propose a Joint Autocorrelation and Cross-correlation (JAC) scheme for estimating the coupled near-field CoA and AoA information. 	 Firstly, based on the dual relationship between digital signal processing (DSP) and array signal processing (ASP), we have identified the dual problem of the near-field problem, namely the Doppler problem.  Therefore, we refer to the near-field problem as the spatial Doppler problem. Inspired by the idea that the frequency change of time-domain signals can be solved by calculating the coherence time, we found that the CoA information of the near-field can be obtained by solving the spatial autocorrelation function of the signal. So, we proposed a scheme of first solving the CoA based on the spatial autocorrelation function of the received signal, and then substituting the obtained CoA to solve the AoA. Due to the fact that the process of solving AoA is similar to the principle of far-field scenarios, which is to find the direction space with the maximum cross-correlation value with the channel, we call this scheme the JAC scheme. Compared with other existing near-field estimation algorithms, the JAC scheme is not affected by AoA when solving CoA. Therefore, we can perform orthogonal decomposition separately in the space of CoA and AoA. This allows some algorithms based on orthogonal decomposition to be transferred to near-field use through the JAC scheme. 
	
	\item
	We proposed the Inverse Sinc Function  (JAC-ISF) algorithm. Since the autocorrelation function of the near-field received signal is in the form of a $sinc$ function, we can use the ${\rm arc} sinc$ function to solve for CoA. However, since the $sinc$ function is not a monotonic function, we need to intercept the values of the $sinc$ function between $(-1,0)$ to form the ${\rm arc} sinc$ function.  The advantage of JAC-ISF algorithm is that it is easy to implement because it does not require iterative or optimization operations. The disadvantage is that we only used the main lobe data of the autocorrelation function and did not use all the data, resulting in a certain degree of information loss and a decrease in algorithm accuracy. 
	
	\item 
	We proposed the Gradient Descent (JAC-GD) algorithm.  In this algorithm, we use the difference between the autocorrelation function and the sinc function as the loss function and employ gradient descent to estimate the value of CoA. Compared to the JAC-ISF algorithm, the JAC-GD algorithm is more complex, but due to its utilization of all autocorrelation function information, it has higher accuracy. 
	
	\item
	We analyzed the time complexity of the JAC scheme and the Cramer-Rao Lower Bound (CRLB) for near-field position estimation. Simulation results show that the JAC-GD algorithm is closer to CRLB compared to other algorithms, and its performance is significantly better than other algorithms at different SNRs and communication distances. This demonstrates that the JAC-GD algorithm achieves better performance while reducing power and time overhead.

\end{itemize}

The remainder of this paper is organized as follows. Section II presents the system model. Section III introduces the JAC scheme, while Section IV propose two near-field channel and position estimation algorithms. We analyze the cramer-rao lower bound (CRLB) and time complexity in Section V. Simulation results are shown in Section VI while  the conclusion is presented in Section VII.

{\itshape Notations}: In this paper, scalars, vectors, and matrices are denoted by italic letters, bold-face lower-case, and upper-case letters, respectively. The space of ${ x} \times { y}$ complex-valued matrices is denoted by $\mathbb{C}^{{x} \times { y}}$. The function $sinc(x)$ is $\frac{\sin(x)}{x}$. For a complex-value vector $\bm x$, ${\bm x}\otimes{\bm y}$ denotes the Kronecker product of $\bm x$ and $\bm y$ while $\Vert{\bm x}\Vert_2$ denotes its  binary norm and diag$({\bm x})$ denotes a diagonal matrix with each diagonal entry being the corresponding entry in $\bm x$. ${\bm x}\cdot{\bm y} $ denotes the dot product between these two vectors, while the cross product between $\bm x$ and $\bm y$ is represented by ${\bm x}{\bm y}$ or ${\bm x}\times {\bm y}$.  ${\bm x}_{i}$ is the  $i$th entry of $\bm x$. For a function ${\bm y}=H({\bm x})$, $H^{-1}({\bm y})$ denotes its inverse function. For a general matrix $\bm A$, ${\bm A}^*$, ${\bm A}^H$, ${\bm A}_{i,j}$ and ${\rm det}(\bm A)$ denote its conjugate, conjugate transpose, the $(i,j)$th entry, and the determinant of the $\bm A$ respectively. $\jmath$ denotes the imaginary unit, i.e., ${\jmath}^2=-1$.



\section{System model }

As shown in Figure 1, we consider the channel model in the uplink scenario. The base station (BS) is a uniform linear array with $N$ antennas located on the $x$-axis, and its reference antenna is located at the origin. The user (UE) is a single antenna located at point ${\bm p}=(p_x,0,p_z)$ on the $xOz$ plane. $\theta$ and $n$ are the elevation angle of the reference antenna towards the user and the index of the antenna, respectively.  The received signal on BS can be written as:
\begin{equation}\label{ini_channelmodel}
	{\bm Y}={\bm h}{\bm s}^H+{\bm N}_{\sigma},
\end{equation}
where $\bm Y \in \mathbb{C}^{N\times T}$, $\bm h \in \mathbb{C}^{N\times 1}$, $\bm s\in \mathbb{C}^{T\times 1}$ and ${\bm N}_{\sigma}\in \mathbb{C}^{N\times T} $ are the  received signal matrix, the channel, the transmitted  signal at $T$ time slot and the  Gaussian white noise with power $\sigma$, respectively. Due to the weak scattering, diffraction, and reflection abilities of high-frequency signals, the channel is often sparse, and its power is dominated by the line of sight (LoS) path \cite{10005200}. Therefore, we only consider the LoS path and treat the non line of sight (NLoS) path as noise. 
The $\bm s$ in equation (\ref{ini_channelmodel}) is:
\begin{equation}
	\bm s=[s(1), s(2), \cdots, s(t),\cdots, s(T)], \;\; 
\end{equation}
where $s(t)$  can be written as:
\begin{equation}
	s(t)=\bm \rho_t e^{\jmath (2\pi f t+\bm \psi _t)},
\end{equation}
where $\bm \rho \in \mathbb{C}^{T\times 1}$, $\bm \psi\in \mathbb{C}^{T\times 1} $ and  $f$ are  the amplitude, phase of the transmitted signal and the carrier frequency, respectively.
For channel $\bm h$, without loss of generality, we use a continuous function $h(x_n)$ to represent it, where $x_n$ is the $x-{\rm axis}$  coordinate of the $n$th antenna. $h(x_n)$ can be written as:
\begin{equation}\label{channel_model}
	\begin{split}
		&h(x_n)=\bm \varrho_n e^{\jmath k\phi(x_n)}\\
		&=\bm \varrho_n e^{-\jmath k\left( \Vert{\bm p}-{\bm x}_n\Vert_2 - \Vert{\bm p} \Vert_2   \right)},
	\end{split}
\end{equation}
where $\bm \varrho\in \mathbb{C}^{N \times 1}$ and $k\phi(x_n)$ are the channel gain and relative delay of the transmitted signal for each subchannel, respectively. $\bm \varrho$ is mainly affected by large-scale fading, while small-scale fading such as near-field effects have little impact on it. Therefore, without loss of generality, we take $\bm \varrho= \bm 1$. In near-field channel estimation and localization problems, we mainly focus on $\phi(x_n)$. For $\Vert{\bm p}-{\bm x}_n\Vert_2$ in $\phi(x_n)$, we use the binomial theorem to expand:
\begin{align}\label{binomial_theorem}
	\begin{split}
		&\Vert {\bm p}-{\bm x}_n \Vert_2\\
		&=\left( r^2 +\left(-2r \sin \theta x_n +x_n^2\right)\right)^{\frac{1}{2}}\\
		&=r+\frac{1}{2}r^{-1}\left(-2r\sin \theta x_n +x_n^2 \right)\\
		&-\frac{1}{8}r^{-3} \left(-2r\sin \theta x_n +x_n^2\right)^2\\
		&+\frac{1}{16}r^{-5} \left(-2r \sin \theta x_n+x_n^2\right)^3+\cdots\\
		&=r -x_n\sin \theta +\frac{x_n^2 \cos^2 \theta}{2r} +\frac{x_n^3\sin \theta \cos^2 \theta}{2r^2}+o(x_n^3),
	\end{split}
\end{align}
where $r=\Vert \bm p \Vert_2$ is the distance between UE and the refrence antenna of the BS.
Based on the relationship with $x_n$, we categorize the terms in equation  (\ref{binomial_theorem}) as constant, linear, quadratic, and higher-order, respectively. Generally, we consider the phase error of $h(x_n)$ negligible when it is less than $\frac{\pi}{8}$. The omitted portion of equation (\ref{binomial_theorem}) represents the infinitesimal of the first few terms; hence, our discussion will focus on these initial terms.

When $\frac{x_n^3 \cos^2\theta \sin\theta}{2r^2}>\frac{\lambda}{16}$, leading to $r <0.62\sqrt{\frac{(Nd)^3}{\lambda}}$ \cite{7942128, DBLP:conf/acssc/BjornsonDS21, DBLP:journals/twc/JiangGJZZ23 }, electromagnetic waves exhibit electrical resistance \cite{DBLP:conf/pimrc/HuIW22} and cannot radiate energy outward. Consequently, in communication scenarios, as long as the signal can propagate, it must satisfy $\frac{x_n^3 \cos^2\theta \sin\theta}{2r^2}<\frac{\lambda}{16}$. Hence, higher-order terms can always be neglected in the communication channel model.

When $\frac{x_n^2 \cos^2\theta}{2r}<\frac{\lambda}{16}$ holds, $r>\frac{2(Nd)^2}{\lambda}$. This condition characterizes the Fraunhofer region, also known as the far-field region, where ${\bm h}[n]$ becomes a function of $\theta$ independent of $r$. Conversely, when $r<\frac{2(Nd)^2}{\lambda}$, the corresponding region is termed the Fresnel or near-field region. 

To sum up,  the phase function of the near-field channel is quadratic, while the far-field channel is a linear function. Thus, the far-field channel and near-field channel can be parameterized as:
\begin{equation}\label{far-field}
	h_{\rm far}(x_n)= e^{\jmath k p_2 x_n}
\end{equation}
and 
\begin{equation}\label{near-field}
	h_{\rm near}(x_n)=e^{\jmath k(p_1 x_n^2+p_2 x_n)},
\end{equation}
respectively. We can observe that the far-field model is a special case of the near-field model when $p_1=0$. Therefore, designing parameter estimation algorithms based on equation (\ref{near-field}) can achieve both far-field channel estimation and near-field channel estimation.

\begin{figure}
	\centering
	\includegraphics[scale=1]{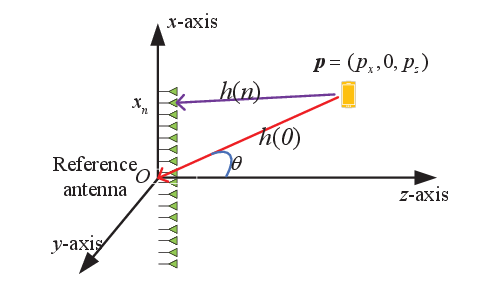}
	\caption{ Near-field channel model for ULA communication system.}     \label{coordinate system}
\end{figure}

\section{JAC Scheme}
\subsection{The Physical Meaning of Near-field Channel Parameters}
In order to find better near-field channel estimation methods, we will analyze the physical meaning of near-field channel parameters in this section.
According to equation (\ref{channel_model}), the UE's location information is contained in $\phi(x)$. According to equations (\ref{binomial_theorem}) and (\ref{near-field}), we can obtain:
\begin{equation}\label{phix}
	\phi(x)=p_1 x^2+p_2 x,
\end{equation}
where
\begin{equation}\label{p1}
	p_1=-\frac{\cos^2 \theta}{2r},
\end{equation}
and
\begin{equation}\label{p2}
	p_2=\sin \theta.
\end{equation}
We can observe that the distance and angle information of the user are coupled in $p_1$, while the information of $p_1$ and $p_2$ are coupled in $\phi$. Therefore, if the distance and angle of the user are estimated directly on $\phi$, there will be a phenomenon of mutual coupling between parameters. 

\begin{figure}
	\centering
	\includegraphics[scale=0.8]{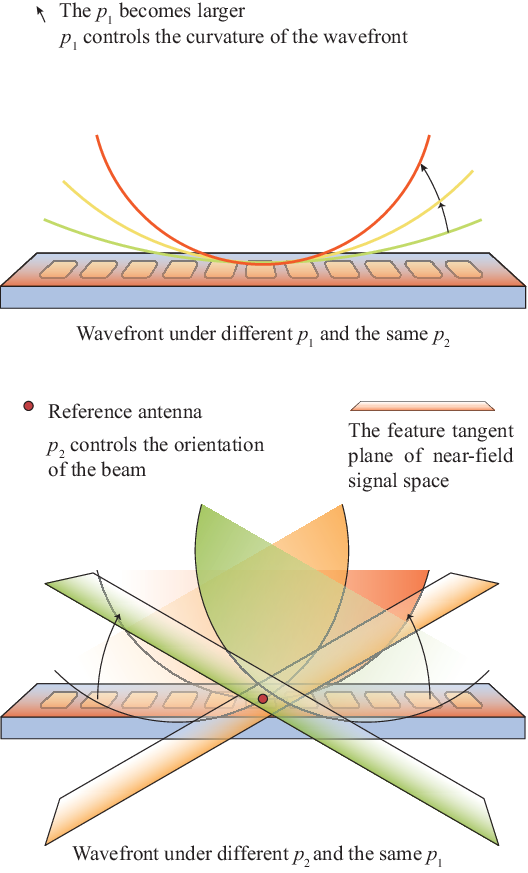}
	\caption{ Parameters for controlling the curvature and direction of spherical waves.}     \label{p1p2}
\end{figure}

We address the parameter coupling phenomenon in near-field channels from two aspects. Firstly, we directly estimate $p_1$ and $p_2$, and then estimate the UE's position using equations ($\ref{p1}$) and ($\ref{p2}$) after obtaining $p_1$ and $p_2$. Secondly, we will find a method to decouple $p_1$ and $p_2$ based on their physical meanings.   We use $\beta(x)$ to represent the direction in which the antenna at position $(x,0,0) $ looks towards the UE, that is:
\begin{equation}
	\beta(x)=\sin \theta_x, \,\,\,\,\,  \beta(x)\in (-1,1),
\end{equation}
where $\theta_x$  is the elevation angle of the antenna at position $(x,0,0)$ towards the UE. According to equation (\ref{phix}), the following holds:
\begin{equation}
	\beta(x)=\frac{{\rm d}\phi(x)}{{\rm d}x}=2p_1x+p_2.
\end{equation}
Since $\beta(0)=p_2$, $p_2$ represents the channel direction at the reference antenna. We define the channel direction at the reference antenna as the angel of arrival (AoA) and the rate of change of a channel with respect to space as the curvature of arrival (CoA) of  the channel, which is:
\begin{equation}\label{CoA}
	{\rm CoA}=\frac{{\rm d}\beta(x)}{{\rm d}x}=2p_1.
\end{equation}
From equation (\ref{CoA}), it can be seen that the physical meaning of $p_1$ is the curvature of the channel. Fig. \ref{p1p2} shows the relationship between $p_1$ and $p_2$ and UE's location. It can be seen that the size of $p_1$ affects the CoA of the wavefront, regardless of its direction while $p_2$ represents the direction of the tangent plane of the wavefront at the reference antenna, without affecting the curvature of the wavefront.
Therefore, as long as we find a method that can estimate channel CoA and AoA separately, we can decouple $p_1$ and $p_2$.

\subsection{The Relationship Between Near-field Problems and Doppler Problems}
In this section, we explore the dual relationship between the near-field problem and the Doppler problem to find a method for estimating the near-field channel CoA, and then use the estimated CoA information to estimate the AoA.

For the time-domain function ${h}(t)=e^{\jmath \phi(t)}$, if $\phi'(t)$ is a constant value, then ${h}(t)$ is stationary, and its frequency spectrum ${H}(\omega)$ does not change with time. If there is Doppler effect in the channel, ${h}(t)$ becomes non-stationary. At this point, $\phi'(t)$ is not a constant value, and ${H}(\omega)$ also changes with time. The coherence time of ${h}(t)$ can reflect the speed at which ${H}(\omega)$ changes over time.

In near-field problems, the channel also exhibits properties similar to the Doppler effect. We use $\kappa=\frac{k_x}{k}$ to represent the direction of the channel, where $k_x$ is the projection of $k$ in the $x$-axis direction.  We select a small interval $(x_0, x_0+\Delta x)$ on the array, where $\phi(x)$ can be seen as a linear function. The pattern on this interval can be written as:
\begin{align}
	\begin{split}
		&H(\kappa)|_{x=x_0}\\
		&=\frac{1}{\Delta x} \left| \int_{x_0}^{x_0+\Delta x}e^{\jmath k \phi(x)}e^{-\jmath k\kappa x}{\rm d} x  \right|\\
		&\frac{1}{\Delta x}\left| \int_{x_0}^{x_0+\Delta x} e^{\jmath k(\phi(x_0)+\phi'(x)(x-x_0))}e^{\jmath k\kappa(x-x_0)}{\rm d} x\right|\\
		&=\frac{1}{\Delta x}\left| \int_0^{\Delta x}e^{\jmath k(\beta(x_0)-\kappa)x}{\rm d}x\right|\\
		&=\frac{1}{\Delta x}\left|\int_0^{\Delta x}e^{\jmath k(\beta(x_0)-\kappa)x}{\rm d} x \right|\\
		&=sinc(\beta(x_0)\frac{k\Delta x}{2}).
	\end{split}
\end{align}
When $\beta(x)$ is a constant independent of $x$, $H(\kappa)$ does not change with the variation of $x$, and at this time $h(x)$ is spatially stationary. When $\beta(x)$ is not a constant, $h(x)$ is spatially non-stationary.

\begin{figure*}
	\centering
	\includegraphics[scale=1]{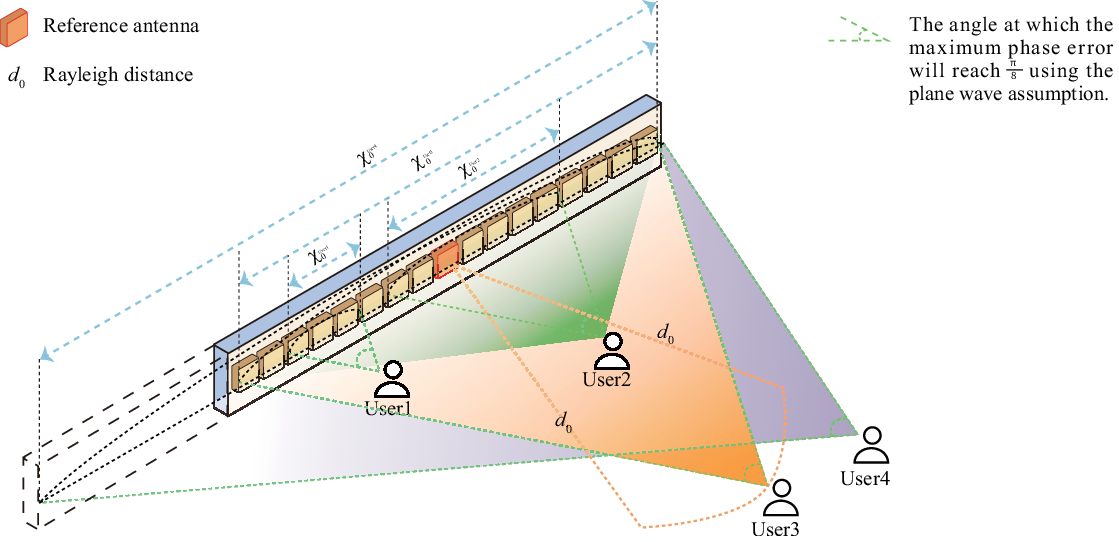}
	\caption{ Spatial Doppler phenomenon.}     \label{spacial_frenquency}
\end{figure*}

We can compare the time signal and spatial signal and find that the Doppler effect causes $h(t)$ to be non-stationary in time, and the near-field effect causes $h(x)$ to be non-stationary in space. Therefore, we call the near-field effect the spatial Doppler effect. Hence, we can imitate the time-domain Doppler estimation method and propose its dual spatial Doppler estimation method.  Thus, we proposed the concept of coherent space by imitating the concept of coherent time.  According to equation (\ref{far-field}), when the distance from the user to the reference antenna is greater than the Rayleigh distance, the channel model can be considered as a far-field model. At this point, $p_1=0$ and $\beta(x) =p_2$ is a constant, so the signal is spatial stationary, meaning that the signal is coherent in the space of the array. Therefore, we define coherent space as the range where the channel can be viewed as a far-field on the line of the array. We use $\chi_0$ to represent the length of the coherent space. As shown in Fig. \ref{spacial_frenquency}, the coherence space size of the channel varies for users at different locations.  For user 3, its distance from the reference antenna is equal to the Rayleigh distance, therefore $\chi_0^{{\rm user} 3}=D$, where $D$ is the physical size of the array. For user 4, it is located in the far-field region of the array, therefore $\chi_0^{{\rm user} 4}>D$. The electromagnetic waves emitted by user 4 can be regarded as plane waves on the array. For near-field users, such as user 1 and user 2, the coherence space of their channels is smaller than the array size, so the channels cannot be considered spatially coherent, i.e. plane wave models cannot be used. And because user 1 is closer to the array than user 2, $\chi_0^{{\rm user} 1}<\chi_0^{{\rm user} 2}$. 

Therefore, we found that by measuring the spatial coherence of the signals received by the array, we can determine the distance from the user to the reference antenna. If the signals on the entire array are coherent, it indicates that the user is in the far field, while if only a portion is coherent, it indicates that the user is in the near field of the array. The smaller the coherence space, the greater the rate of change of channel direction with space, that is, the larger the CoA. So we can solve $p_1$  by doing autocorrelation on the signal. In the next subsection, we will provide a detailed introduction to the functional relationship between signal autocorrelation and $p_1$.

\subsection{Estimation Model of Near-field Channel}
We define the amplitude spectrum of the spatial autocorrelation function of the received signal as:

\begin{align}\label{cchi}
	\begin{split}
		&c \left( \chi \right) =\frac{1}{D-\widetilde{D}} \left| \int_{\widetilde{D}}^{D}h \left( x \right) h^* \left( x-\chi \right) {\rm d}x\right|\\
		&=\frac{1}{D-\widetilde{D}} \left|\int_{\widetilde{D}}^D e^{\jmath k \left( p_1 x^2+p_2 x \right) }e^{-\jmath k \left( p_1 \left( x-\chi \right) ^2+p_2 \left( x-\chi \right)  \right) }{\rm d} x\right|\\
		&=\frac{1}{D-\widetilde{D}} \left|\int_{\widetilde{D}}^D e^{\jmath k 2p_1 \chi x}e^{\jmath k \left( -p_1\chi^2+p_2 \chi \right) }{\rm d} x\right|\\
		&=\frac{1}{D-\widetilde{D}} \left|\int_{\widetilde{D}}^D e^{\jmath k 2p_1 \chi x}{\rm d} x\right|\\
		&=\left| sinc \left( kp_1\chi \left( D-\widetilde{D} \right)  \right) \right|,
	\end{split}
\end{align}
where $\chi$ represents the displacement in the autocorrelation function.
According to equation (\ref{cchi}), we can write the solution of $p_1$ as:
\begin{equation}\label{p1_estimate}
	p_1=\min_{p_1} \left|c(\chi) -\left| sinc\left(kp_1 \chi\left(D-\widetilde{D} \right) \right)\right|\right|.
\end{equation}
After estimating the parameter $p_1$, we can construct a new signal form to transform the near-field problem into a far-field problem:
\begin{equation}\label{new_signal}
	\widetilde{h}(x)=e^{\jmath k \hat{p}_1 x^2}h(x).
\end{equation}
At this point, $\widetilde{h}(x)$ is equivalent to a far-field channel, so we can use traditional far-field channel estimation methods to estimate parameter $p_2$. Finally, based on equation (\ref{near-field}), we can obtain the estimated channel $\hat{h}$, and based on equations (\ref{p1}) and (\ref{p2}), we can obtain $\hat{\theta}$ and $\hat{r}$.

In summary, we decompose the near-field  estimation process into two steps: estimating $p_1$ and $p_2$. The process of estimating $p_1$ is an estimation method similar to the Doppler effect, which uses autocorrelation function for estimation. The process of estimating $p_2$ is essentially an estimation of the directional pattern, which is to determine which direction the signal has the greatest cross-correlation with. Therefore, the estimation model proposed in this paper is named the joint auto-correlation and cross-correlation (JAC) scheme.

\section{Near Field Localization Algorithms Based on JAC Scheme}
The JAC model is mainly divided into two stages, with the second stage essentially being far-field AoA estimation. Therefore, in this section, we propose two algorithms to solve the CoA estimation problem in the first stage of the JAC scheme.

\subsection{Inverse sinc function algorithm}
In this part, we use the inverse function of the sinc function to solve the parameter $p_1$, so we call this algorithm the JAC-ISF algorithm.

\begin{figure} 
	\centering
	\includegraphics[width=\linewidth]{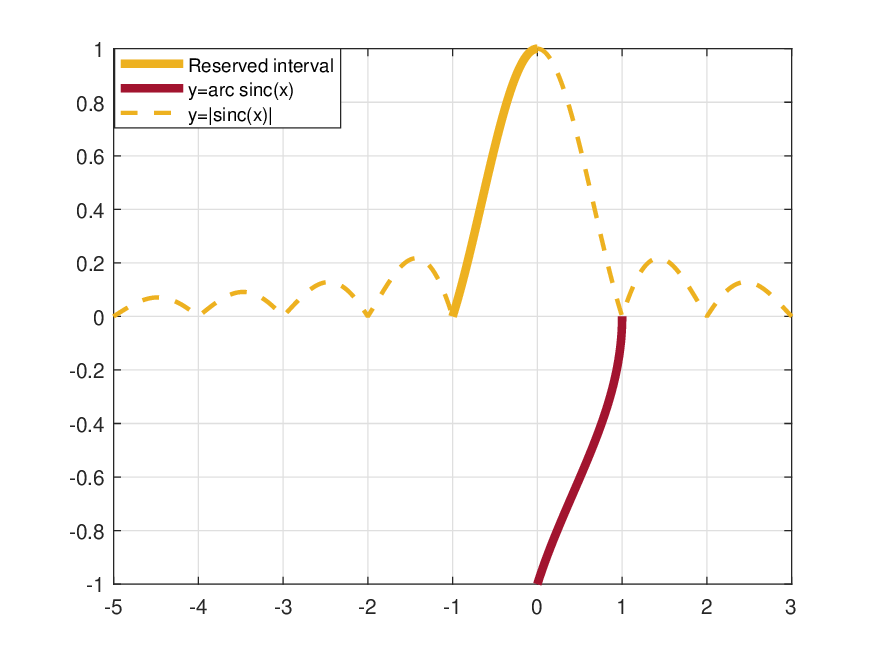}
	\caption{The ${\rm arc}sinc$ function.}   \label{arcsinc} 
\end{figure}

According to equation (\ref{cchi}), the analytical solution of $p_1$ can be expressed as the inverse function of the $sinc$ function. However, the $sinc$ function is not a monotonic function and does not have an inverse function.  Therefore, we need to truncate the $sinc$ function and use its monotonic interval for channel estimation.  Firstly, we can discard the sidelobes and only retain the main lobe. This is because near-field scenes are often accompanied by the presence of XL-MIMO. In XL-MIMO scenario, the sidelobe amplitude of the $sinc$ function is much lower than that of the main lobe, so the anti noise performance of the sidelobes is worse.  Secondly, according to equation (\ref{p1}), we know that $p_1$ must be a negative value. Therefore, we only need to retain the part of the $sinc$ function $(-1,0)$. On this interval, the $sinc$ function is a monotonically increasing function and therefore has an inverse function.  Thus, ${\rm arc}sinc(x)$ can be written as:
\begin{equation}
	{\rm arc}sinc(x)=\{ y|x=sinc(y), y\in(-1,0)\}.
\end{equation}
Fig. \ref{arcsinc} shows the graph of the ${\rm arc}sinc$ function.

Since we need to compare the autocorrelation function with the $sinc$ function to obtain the value of $p_1$, the autocorrelation function also needs to capture points within the main lobe. According to equation (\ref{cchi}), the estimated value of the autocorrelation function can be written as:
\begin{equation}\label{ceta}
	\hat{c}[\eta]=\frac{1}{T}\frac{1}{N-\xi}\sum_{t=1}^T\sum_{n=\xi+1}^{N} {\bm Y}[n,t]{\bm Y}^*[n-\eta,t],\,\,\, \eta\in [1,\xi],
\end{equation}
where $\xi \in \mathbb{Z}$ and $\eta$ are the number of points in the autocorrelation function and  the number of shifts when performing autocorrelation functions.  We can see that when the value of $\xi$ is small, the estimated value of the autocorrelation function is more accurate. However, the number of points in the autocorrelation function decreases, resulting in insufficient data for estimating $p_1$. On the contrary, if the value of $\xi$ is large, it will lead to inaccurate estimation of the autocorrelation function.  In practical applications, if there is prior information, the appropriate $\xi$ should be selected based on the prior information. In order to consider a more general situation, we choose $\xi=\lfloor\frac{N}{2}\rfloor$. Then, we need to determine whether the autocorrelation function has zero points to determine if all data is within the main lobe.  Due to the fact that autocorrelation functions may not necessarily have zero points when $\eta$ is an integer, we cannot extract data within the main lobe by finding $\eta$ values that satisfy $c[\eta]=0$. Therefore, we select a small value\footnote{Here we have chosen $\delta=0.1$ based on experience. In fact, the choice of $\delta$ does not have a significant impact on the performance of the algorithm, so it is sufficient to choose the appropriate one when applying this algorithm.} $\delta=0.1$  and use $\Upsilon$ to represent the set of $\eta$ that satisfies $c[\eta]\le \delta$, that is:
\begin{equation}\label{Uplison}
	\Upsilon=\{\eta|c[\eta]\le \delta\}.
\end{equation}
Thus, the number of points retained for the autocorrelation function is:
\begin{equation}\label{Neta}
	N_{\eta}= \left\{
	\begin{aligned}
		\min{\Upsilon}, \,\,\, \Upsilon \neq \emptyset, \\
		\xi,  \,\,\, \Upsilon = \emptyset. 
	\end{aligned}
	\right.
\end{equation}
Substituting $\chi=\eta d$, according to equations (\ref{cchi}) and (\ref{p1_estimate}), the estimated value of $p_1$ is:
\begin{equation}\label{p1hat}
	\hat{p}_1=\frac{1}{N_\eta}\sum_{\eta=1}^{N_\eta}\frac{{\rm arc}sinc(\hat{c}[\eta])}{k\eta d^2 (N-N_\eta)}.
\end{equation}
According to equation (\ref{new_signal}), we use the estimated $p_1$ to construct an equivalent far-field signal:
\begin{equation}\label{equivalent_far}
	\widetilde{\bm Y}[n,t]=e^{\jmath k p_1 (nd)^2}{\bm Y}[n,t].
\end{equation}
Then  we use the MUltiple SIgnal Classificatio (MUSIC) algorithm on $\widetilde{\bm Y}$ to estimate the direction of the maximum eigenvalue, which is $p_2$.
Finally, the estimated channel is:
\begin{equation}\label{estimated_channel}
	\hat{\bm h}[n]=e^{\jmath k (\hat{p}_1 n^2 d^2+\hat{p}_2 nd)}.
\end{equation}
Since this algorithm obtains $p_1$ by inverting the $sinc$ function, we refer to it as the Inverse Sinc Function  (ISF) algorithm.
The details of this algorithm are summarized in Algorithm 1.

\begin{algorithm}
	\caption{JAC-ISF algorithm}
	\begin{algorithmic}[1]\label{alg1}
		\STATE {Require: $\bm Y$, $N$, $T$, $\xi$.}
		\STATE{Estimate $c[\eta]$ based on equation (\ref{ceta}).}
		\STATE{Determine $\Upsilon$  based on equations (\ref{Uplison}).}
		\STATE{According to equation (\ref{Neta}), determine the number of data $N_\eta$ used to estimate $p_1$.}
		\STATE{According to equation (\ref{p1hat}), obtain the estimated value $\hat{p}_1$ of $p_1$.}
		\STATE{According to equation (\ref{equivalent_far}), form the equivalent far-field signal $\widetilde{\bm Y}$.}
		\STATE{Estimate parameter $p_2$  using the MUSIC algorithm.}
		\STATE{ Get $\hat{h}$   according to equation (\ref{estimated_channel}).}
		\STATE{Get $\hat{\theta}$  and $\hat{r}$ according to equations (\ref{p1}) and (\ref{p2}).   }
	\end{algorithmic}
\end{algorithm}
\vspace{-0in}%

\subsection{Gradient descent method for calculating CoA}
In this subsection, we use gradient descent to solve the parameter $p_1$. Therefore, we call this algorithm the JAC-GD algorithm.

Similar to the JAC-ISF algorithm, we obtain an estimate of the autocorrelation function using equation (\ref{ceta}). According to equation (\ref{p1_estimate}), we set the loss function as:
\begin{equation}\label{Loss}
	Loss=\sum_{\eta=1}^\xi\left| \hat{c}[\eta]-\left|sinc(kp_1\eta d^2 (N-\xi))\right| \right|.
\end{equation}
The gradient of the Loss function is:
\begin{equation}\label{grad}
	grad=\frac{{\rm d}Loss}{{\rm d}p_1}.
\end{equation}
We update $p_1$ based on the learning rate $\alpha$, that is:
\begin{equation}\label{p1_itr}
	p_1(n_{\rm itr})=p_1(n_{\rm itr}-1)-\alpha(n_{\rm itr})grad,
\end{equation}
where $n_{\rm itr}$ is the index of the number of iterations. We adopt the learning rate of inverse time decay, that is:
\begin{equation}\label{alpha}
	\alpha(n_{\rm itr})=\alpha_0 \frac{1}{1+\gamma(n_{\rm itr}-1)},
\end{equation}
where $\alpha_0$ is the initial learning rate and $\gamma$ is the decay factor.
The details of this algorithm are summarized in Algorithm 2.

\begin{algorithm}
	\caption{JAC-GD algorithm}
	\begin{algorithmic}[1]\label{alg2}
		\STATE {Require: $\bm Y$, $N$, $T$, $\alpha_0$, $\gamma$, $N_{\rm itr}$ and $\xi$}
		\STATE{Initialize: $P_1=0$.}
		\STATE{Obtain the  $\hat{c}[\eta]$ using equation (\ref{ceta}).}
		\FOR{$n_{\rm itr}=1:N_{\rm itr}$}
		\STATE{Obtain the loss function through equation (\ref{Loss}).}
		\STATE{Calculate the current gradient as equation (\ref{grad}).}
		\STATE{Update $\alpha$ according to equation (\ref{alpha}).}
		\STATE{Update $p_1$ according to equation (\ref{p1_itr}).}
		\ENDFOR
		\STATE{Obtain the equivalent far-field channel $\tilde{\bm h}$ according to equation (\ref{equivalent_far}).}
		\STATE{Estimate $p_2$ using MUSIC algorithm.}
		\STATE{Obtain the estimated channel $\hat{\bm h}$ based on equation (\ref{estimated_channel}).}
		\STATE{Obtain the polar coordinates of the UE's position based on equations (\ref{p1}) and (\ref{p2}).}
		
	\end{algorithmic}
\end{algorithm}
\vspace{-0in}%

\section{Cramer's lower bound and time complexity analysis}

\subsection{Analysis of Cramer's Lower bound for Angle and Distance}
We define the unknown parameter vector as:
\begin{equation}
	\epsilon_1=[p_1, p_2, {\bm \psi}, {\bm \rho}, \sigma^2 ]
\end{equation}
The probability density function of the received signal is:
\begin{equation}
	p({\bm Y}|\epsilon_1)=\frac{1}{\pi^{NT}{\rm det}(\bm R)} e^{-({\bm Y}-{\bm \mu}^H) {\bm R}^{-1}({\bm Y}-{\bm \mu})},
\end{equation}
where ${\bm R = \sigma^2 {\bm I}_{NL}}$ and 
\begin{equation}
	\begin{split}
		&{\bm \mu}=[s(1)h_{\rm near}^{\rm T}(p_1,p_2),s(2)h_{\rm near}^{\rm T}(p_1,p_2),\\
		&\cdots, s(T)h_{\rm near}^{\rm T}(p_1,p_2)].
	\end{split}
\end{equation}
We define $\mathbb{E}\{(\hat{\bm \epsilon}_1-{\bm \epsilon}_1)(\hat{\bm \epsilon}_1-{\bm \epsilon}_1)^{\rm T}  \}$ as the covariance matrix of the unknown parameter vector $\bm \epsilon$, where $\hat{\bm \epsilon}$ is the unbiased estimator of $\bm \epsilon$. The covariance matrix has the following inequality relationship:
\begin{equation}
	{\rm MSE}([{\bm \epsilon}_1]_i)=\mathbb{E}\{([\hat{\bm \epsilon}_1]_i-[{\bm \epsilon}_1]_i)^2   \} \geq {\rm CRLB}([{\bm \epsilon}_1]_i),
\end{equation}
where the  $\rm CRLB$ is cramer-rao lower bound (CRLB) and satisfies:
\begin{equation}
	{\rm CRLB}([{\bm \epsilon}_1]_i)=[{\rm FIM}^{-1}({\bm \epsilon}_1)]_{i,i}.
\end{equation}
The Fisher information matrix ${\rm FIM}^{-1}({\bm \epsilon}_1)$ can be obtained by the Slepian-Bangs equation \cite{DBLP:journals/tsp/Collier05}, which is shown in (\ref{FIM}):
\begin{align}\label{FIM}
	\begin{split}
		[{\rm FIM}({\bm \epsilon}_1)]_{i,i}&=2 \mathcal{R}\{\frac{\partial \mu^{\rm H}}{\partial [{\bm \epsilon}_1]_i}{\bm R}^{-1} \frac{\partial \mu}{\partial [{\bm \epsilon}_1]_j}   \} \\
		&+{\rm Tr}\{{\bm R}^{-1} \frac{\partial {\bm R}}{\partial [{\bm \epsilon}_1]_i}  {\bm R}^{-1} \frac{\partial {\bm R}}{\partial [{\bm \epsilon}_1]_j}     \}.
	\end{split}
\end{align}
When $T$ is large, the dimensionality of the FIMs also becomes large, making the inverse operation more difficult.  Therefore, we solve ${\rm FIM}^{-1}$ by solving the Schur complement of the matrix.  After simplification, we can obtain the CRLB of unknown parameters $p_1$ and $p_2$:
\begin{equation}
	\begin{split}
		&{\rm CRLB}(p_1)=[{\rm FIM}^{-1}({\bm \epsilon}_1)]_{1,1}=\frac{\sigma^2}{2k^2}\frac{f(0)}{\Vert {\bm \rho}\Vert_2^2}\\
		&\frac{ f(0)f(2)-f^2(2)  }{ (f(0)f(2)-f^2(1)) (f(0)f(4)-f^2(2)   )( f(0)f(3)-f(1)f(2) )^2   },
	\end{split}
\end{equation}
\begin{equation}
	\begin{split}
		&{\rm CRLB}(p_2)=[{\rm FIM}^{-1}({\bm \epsilon}_1)]_{2,2}=\frac{\sigma^2}{2k^2}\frac{f(0)}{\Vert {\bm \rho}\Vert_2^2}\\
		&\frac{ f(0)f(4)-f^2(2)  }{ (f(0)f(2)-f^2(1)) (f(0)f(4)-f^2(2)   )( f(0)f(3)-f(1)f(2) )^2   },
	\end{split}
\end{equation}
where $f(x)=\sum_{n=1}^N (nd)^x$.

For the localization problem, we want to know the CRLB of $\theta$ and $r$, so we define the parameter vector $\bm \epsilon_2$ as:
\begin{equation}
	\bm \epsilon_2=g(\bm \epsilon_1)=[\theta, r, \bm \psi^{\rm T}, \bm \rho^{\rm T}].
\end{equation}
We exploit the theory of parameter transformation for CRLB of $\bm \epsilon$ \cite{1994Fundamentals}:
\begin{equation}
	{\rm CRLB}(\bm \epsilon_2)=\frac{\partial g(\bm \epsilon_1)}{\bm \epsilon_1}{\rm CRLB}(\bm \epsilon_1)(\frac{\partial g(\bm \epsilon_1)}{\bm \epsilon_1})^{\rm T}.
\end{equation}
By combining equations (\ref{p1}) and (\ref{p2}), we can obtain the CRLB of $\theta$ and $r$ as follows:
\begin{equation}
	\begin{split}
		&{\rm CRLB}(\theta)=\frac{\sigma^2}{2}\frac{1}{k^2 d^2 \cos^2(\theta)}\frac{f(0)}{\Vert \bm \rho \Vert_2^2}\\
		&\frac{ f(0)f(4)-f^2(2)  }{ (f(0)f(2)-f^2(1)) (f(0)f(4)-f^2(2)   )( f(0)f(3)-f(1)f(2) )^2   },
	\end{split}
\end{equation}
and (\ref{CRLBr}).

According to equation (\ref{CRLBr}), we can see that the larger $r$, the higher the $\rm CRLB$, and when $r \rightarrow \infty $, the ${\rm CRLB}(r)\rightarrow \infty$. This is consistent with the conclusion that the array cannot determine the distance from the user to the array in far-field situations.

\begin{figure*}[!t]
	\normalsize
	\setcounter{MYtempeqncnt}{\value{equation}}
	\begin{equation}\label{CRLBr}
		{\rm CRLB}(r)=2 \sigma^2 \frac{f(0)}{\Vert \bm \rho \Vert_2^2} \frac{r^2}{k^2 d^4 \cos^4 \theta}
		\frac{ d^2 \sin^2\theta (f(0)f(4)-f^2(2))-2rd\sin \theta(f(1)f(2)-f(0)f(3))+r^2(f(0)f(2)-f^2(1)) }{ (f(0)f(2)-f^2(1)) (f(0)f(4)-f^2(2))-(f(0)f(3)-f(1)f(2))^2 }.
	\end{equation}
	\hrulefill
	\vspace*{4pt}
\end{figure*}

\subsection{Time complexity analysis of JAC-ISF algorithm and JAC-GD algorithm}
The computational complexities of the proposed ISF algorithm is analyzed, which
are summarized in Table I. For solving and  the autocorrelation function, i.e. equation (\ref{ceta}), the time complexity is $\mathcal{O}(TN)$.     The time complexity of the process of extracting the autocorrelation function in equations (\ref{Uplison}) and (\ref{Neta}) are $\mathcal{O}(N)$.    The time complexity of using autocorrelation function to estimate $p_1$ in equation (\ref{p1hat}) is $\mathcal{O}(N_\eta)$, but $sinc$ $N_\eta$ is proportional to $N$, the time complexity is $\mathcal{O}(N)$. In summary, the total time complexity of estimateing $p_1$ is $\mathcal{O}(TN)$. The process of estimating $p_2$ is the MUSIC algorithm, so the time complexity is $\mathcal{O}(TN)$. Finally, the total time complexity of the ISF algorithm is $\mathcal{O}(TN)$. 

In the P-SOWP algorithm, $S$ refers to the number of grids divided by distance. Although \cite{DBLP:journals/tcom/CuiD22} mentions that $S$ is not a  large number compared to $N$. However, as $N$ increases, $S$ also increases synchronously. This is because when $N$ increases, the distance resolution of the array increases.  In the time complexity of the P-SIGW algorithm, $N_{\rm iter}$ refers to the number of iterations of the algorithm.

Through the comparison of time complexity in Table I, we found that the time complexity of most near-field estimation algorithms is the time complexity of distance estimation multiplied by the time complexity of angle estimation \cite{9802728}.  This is because most algorithms use the principle of cross-correlation to estimate the near-field channel. The principle of cross-correlation refers to determining which space a user is in by examining the correlation between channels and different spaces.  This type of channel estimation algorithm based on cross-correlation relies on orthogonal spatial decomposition. In near-field problems, it is difficult to divide the entire space into different orthogonal spaces due to the coupling of distance and direction variables. Moreover, due to the addition of distance as a dimension for measuring space, the time complexity of near-field algorithms exhibits multiplicative time complexity. The JAC model and ISF algorithm proposed in this paper decouple two parameters in near-field channel estimation, and make the time complexity of the entire algorithm exhibit additive time complexity in terms of angle and distance. 

\begin{table}[!htbp]\label{time_complexity}
	\centering
	\caption{\textsc{Time Complexity of Near Field Algorithms}}
	\begin{tabularx}{0.33\textwidth}{|c|c|}
		\hline
		\textbf{ Algorithm name }  & \textbf{Time complexity  }\\
		\hline
		{Proposed ISF } & {$\mathcal{O}(TN)$ }\\
		\hline
		{P-SOWP }  & {$\mathcal{O}(TNS)$ }\\
		\hline
		{P-SIGW }  & {$\mathcal{O}(TNS)+\mathcal{O}(N_{\rm iter}T^2)$ }\\

		\hline
	\end{tabularx}%
	\label{t2}%
	
\end{table}%

\section{Simulation Results}
In this section, numerical results are provided to validate the effectiveness of the proposed algorithm. In our simulations, the BS is a linear array with $N=200$. The carrier frequency is 30GHz. Under this setting, the Rayleigh distance of BS is $200$m. In common cellular scenarios, most users are located within the near-field range of this scenario.  Due to the poor scattering, diffraction, and reflection ability of high-frequency signals, we only use information from the LoS path for channel estimation, and the NLoS path is considered as the noise component. 

\subsection{The Results of Channel Estimation}
We use two metrics to measure the performance of the proposed ISF algorithm, which are achievable rate and normalized minimum mean square error (NMSE) criterion. The achievable rate refers to the information rate that the BS can receive by utilizing the estimated channel for beamforming,  that is:
\begin{equation}\label{achi_rate}
	{\rm Rate}=log_2(1+\frac{P_s N}{\sigma^2}\frac{\vert {\bm h}^H \hat{\bm h}\vert^2}{\Vert { \hat{\bm h}} \Vert_2^2}),
\end{equation}
where $P_s$ is the power of source signal.
If the Channel State Information (CSI) is perfect, the achievable rate is:
\begin{equation}
	{\rm Rate}_{\rm max}=log_2(1+\frac{P_s N}{\sigma^2}).
\end{equation}
The NMSE is:
\begin{equation}
	{\rm NMSE}=\mathbb{E}(\frac{\Vert {\bm h}-\hat{\bm h}\Vert_2^2}{\Vert {\bm h}\Vert_2^2}).
\end{equation}

For a comparative analysis, the proposed codebook schemes are compared with the following schemes:

\begin{itemize}
	\item SWOMP: It is an angle domain on-grid algorithm \cite{8306126}. It only searches for angles and not distances, making it suitable for far-field scenario rather than near-field.
	
	\item SS-SIGW-OLS: It is an angle domain off-grid algorithm \cite{9246294}, which is only applicable to far-field search like the SWOMP algorithm. Due to its lack of grid quantization errors, its algorithm has higher precision.
	
	\item P-SOWP: It is an on-grid algorithm in polar domain \cite{DBLP:journals/tcom/CuiD22}. This algorithm divides the near-field space into different grids based on angle and distance, and then searches for which grid the UE is in. The angle division method is the same as the far-field, which is uniformly divided. The division method in terms of distance is non-uniform, with denser grids at closer distances.
	
	\item P-SIGW: This is an off-grid algorithm in the polar domain \cite{DBLP:journals/tcom/CuiD22}, which is based on the P-SOWP algorithm and uses gradient descent to continue searching within the grid, ultimately determining the location of the UE.
	
	\item MUSIC: This is a classic algorithm for far-field spatial spectrum estimation. Because in both JAC-ISF and JAC-GD algorithms, the estimation of $p_2$ is based on the MUSIC algorithm, we use the MUSIC algorithm  as a comparison scheme to demonstrate to what extent the estimation of $p_1$ in the JAC scheme compensates for the error in estimating the near-field using the MUSIC algorithm.
	
\end{itemize}

\begin{figure} 
	\centering
	\includegraphics[width=\linewidth]{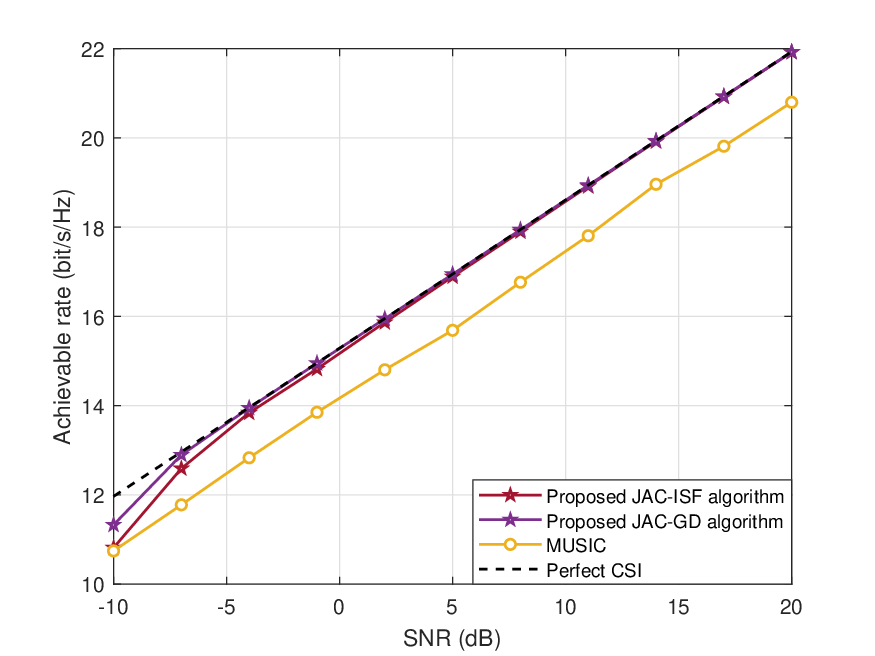}
	\caption{ Comparison of achievable rates of algorithms under different SNR with $T=32$ and $|{\bm p}|\in (10, 50)$.}   \label{achievablerate_snr} 
\end{figure}

Fig. \ref{achievablerate_snr} shows the comparison of achievable rates between proposed  JAC-ISF, JAC-GD algorithms and MUSIC algorithm at different SNRs. All algorithms are performed with a snapshot points of 32. In this simulation experiment, the UE is located on the $z$-axis at a distance of 10 m to 50 m, which belonging to a deeper near-field range. We can see that when SNR=-10 dB, due to excessive noise, JAC-ISF cannot accurately estimate $p_1$, resulting in its performance being similar to the MUSIC algorithm. This indicates that the estimation of $p_1$ is close to complete failure at -10 dB.   However, as the SNR increases, the JAC-ISF algorithm begins to significantly outperform the MUSIC algorithm. Especially when the SNR exceeds 5 dB, the performance of the JAC-ISF algorithm almost conforms to the bound curve. This experiment proves that, except for a few extreme scenarios, the JAC-ISF algorithm is effective in most cases.  For JAC-GD, we can see that its performance is also better than the MUSIC algorithm at -10 dB, indicating that JAC-GD is still effective at low signal-to-noise ratios. And JAC-GD algorithm is closer to the solution under perfect CSI than JAC-ISF algorithm, which indicates the excellent performance of JAC-GD. 

\begin{figure}[!t]
	\centering
	\subfloat[ $T=32$]{
		\includegraphics[width=\linewidth]{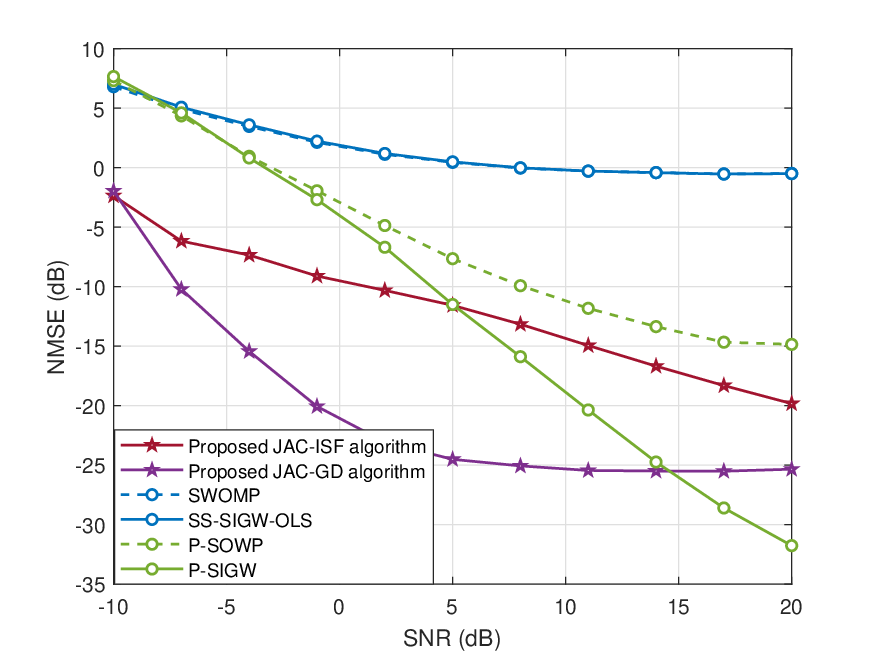}}\\
	\subfloat[ $T=8$]{
		\includegraphics[width=\linewidth]{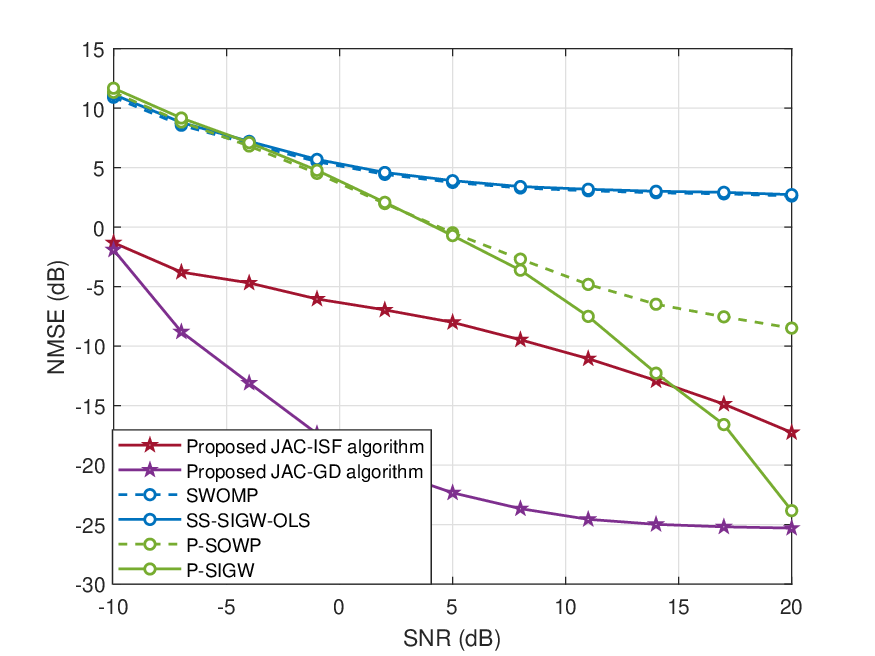}}
	\caption{Comparison of NMSE performance between ISF algorithm and other algorithms under different SNRs.}
	\label{NMSE_snr}
\end{figure}

Fig. \ref{NMSE_snr} shows the NMSE performance  under different SNRs.  The Fig. \ref{NMSE_snr} (a) shows the curve for a snapshot points of 32. We can see that the proposed JAC-ISF and JAC-GD algorithms perform significantly better than other algorithms in the low SNR regime, but their performance limits are not as good as P-SIGW in the high SNR regime. This indicates that the JAC-ISF and JAC-GD algorithms have strong anti noise performance and are more stable in the presence of high noise levels.  
In the high SNR regime, the JAC-GD and JAC-ISF algorithms perform better than the SWOMP, SS-SIGW-OLS, and P-SOWP algorithms, but worse than the  P-SIGW algorithm. This indicates that the upper performance limits of the JAC-ISF and JAC-GD algorithms are not as good as the P-SIGW algorithm, but in the SNR regime of normal communication scenarios, the JAC-ISF and JAC-GD algorithms perform better. 	
Fig. \ref{NMSE_snr} (b) shows the algorithm performance when the $T$ is 8. We found that compared to Fig. \ref{NMSE_snr} (a), the P-SIGW algorithm requires  higher SNRs to exceed the performance of JAC-ISF and JAC-GD. At this point, only when the SNR reaches 14 dB and 20 dB or above, can the performance of the p-algorithm surpass that of JAC-ISF and JAC-GD, respectively. In addition, we found that the performance curves of JAC-ISF and JAC-GD are not significantly different in Fig. \ref{NMSE_snr} (a) and (b). This indicates that the performance of JAC-ISF and JAC-GD algorithms is less affected by the number of snapshots, and subsequent results will also demonstrate this.

\begin{figure} 
	\centering
	\includegraphics[width=\linewidth]{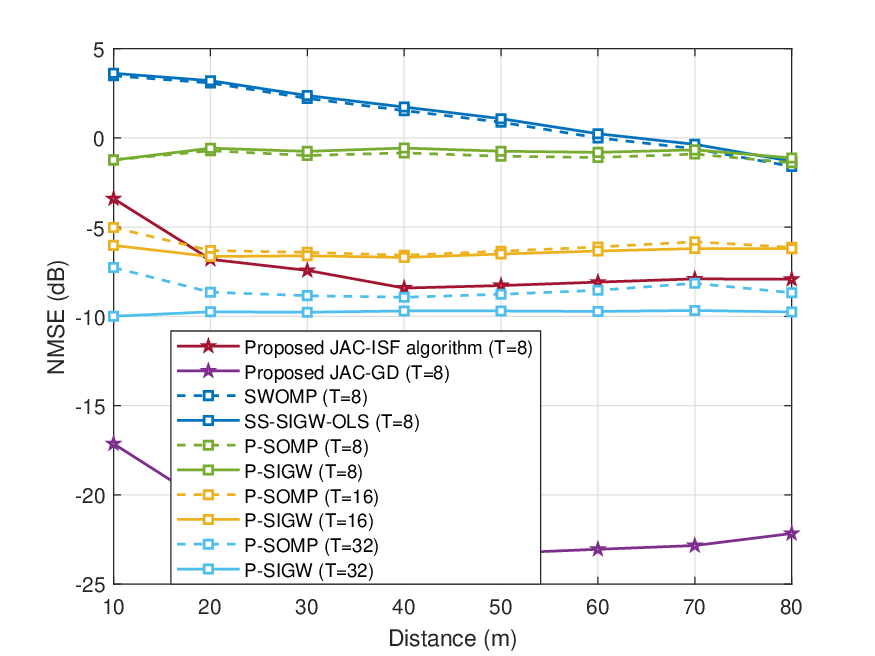}
	\caption{ Comparison of NMSE at different distances with ${\rm SNR}=5$.}   \label{distance_NMSE} 
\end{figure}

Fig. \ref{distance_NMSE} shows the NMSE performance vers distances.  We can see that under the same snapshot points, the performance of the JAC-ISF algorithm is significantly better than other algorithms except the JAC-GD. We can find that when the communication distance is greater than 20 m, the JAC-ISF algorithm can save twice the number of snapshots compared to the P-SIGW algorithm. This indicates that the JAC-ISF algorithm is not only easy to implement, but also has low time complexity. For the JAC-GD algorithm, we can see that its performance at $T=8$ is even better than that of P-SIGW at $T=32$. This demonstrates the significant advantage of the JAC-GD algorithm in the low SNR regime.

\begin{figure} 
	\centering
	\includegraphics[width=\linewidth]{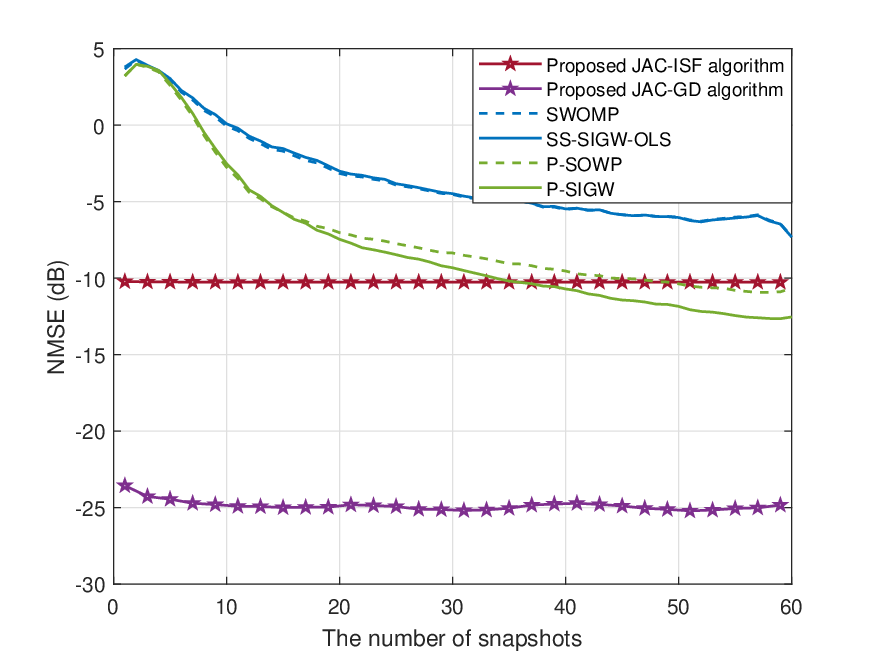}
	\caption{Comparison of achievable rate at different snapshot points with $\rm SNR=5$.}   \label{plength} 
\end{figure}

Fig. \ref{plength} shows the performance comparison of various algorithms under different snapshot points, with the UE distance set to 10 m to 100 m. As illustrated in the Fig. \ref{plength}, the performance of the JAC-ISF and JAC-GD algorithms remain basically unchanged as the number of snapshots increases. This is because in XL-MIMO, the number of spatial samples is relatively large, so even with just one snapshot, the spatial autocorrelation function can be accurately estimated, thereby accurately estimating the channel. 

\subsection{The Results of Position Estimation}
In the problem of location estimation, we use the Root Mean Squared Error (RMSE) criterion to measure algorithm performance, which can be written as:
\begin{equation}
	{\rm RMSE}=\Vert{\bm p}-{\hat{\bm p}}\Vert_2^2,
\end{equation}
where the ${\hat{\bm p}}$ is the estimated UE's position, which can be obtained using $\hat{\theta}$ and $\hat{r}$.

\begin{figure} 
	\centering
	\includegraphics[width=\linewidth]{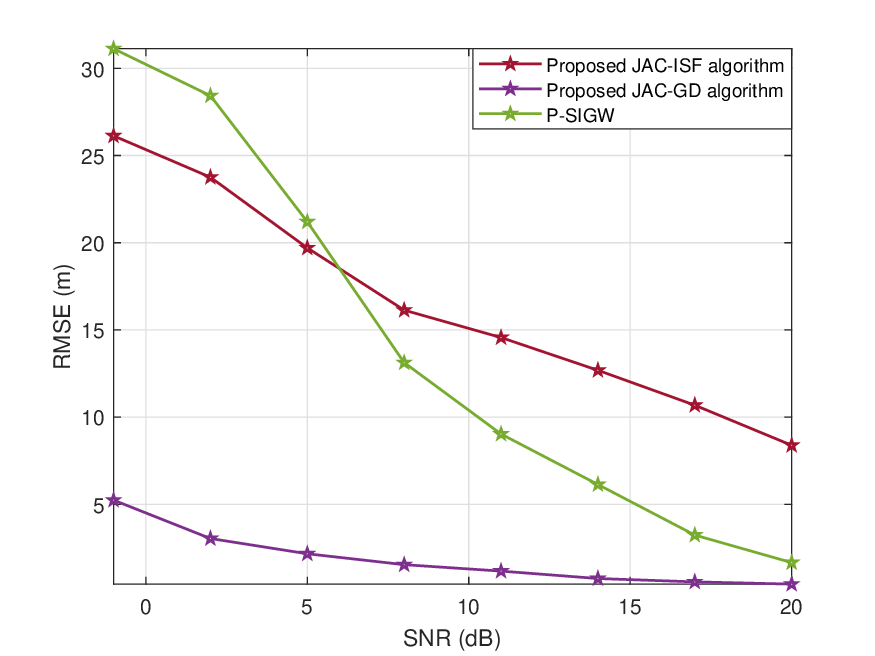}
	\caption{ Comparison of RMSE at different SNRs with $T=8$.}   \label{snr_RMSE} 
\end{figure}

\begin{figure} 
	\centering
	\includegraphics[width=\linewidth]{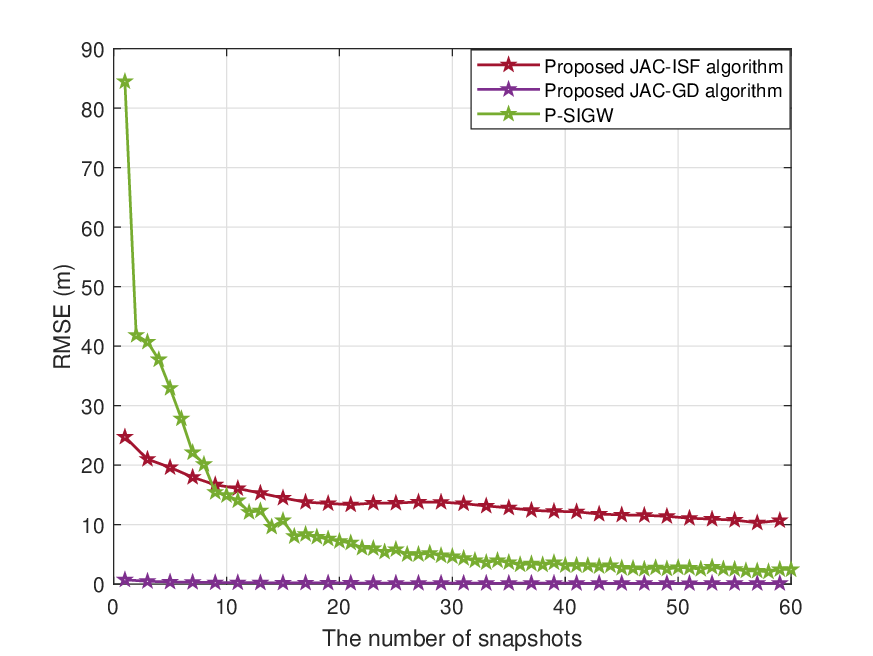}
	\caption{ Comparison of RMSE  with different snapshot points with ${\rm SNR}=5$.}   \label{plength_RMSE} 
\end{figure}

\begin{figure}[!t]
	\centering
	\subfloat[$\theta$]{
		\includegraphics[width=\linewidth]{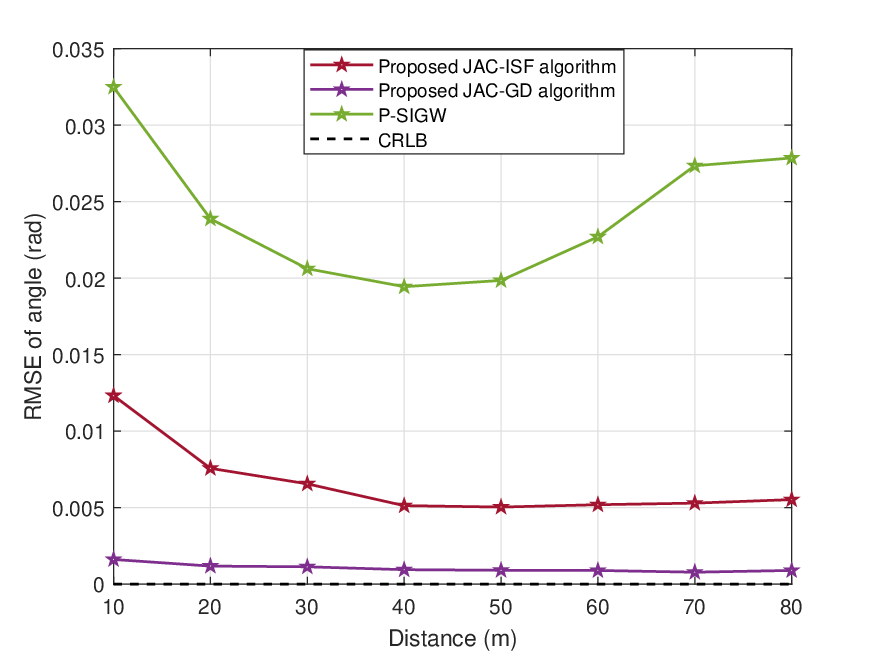}}\\
	\subfloat[ $r$]{
		\includegraphics[width=\linewidth]{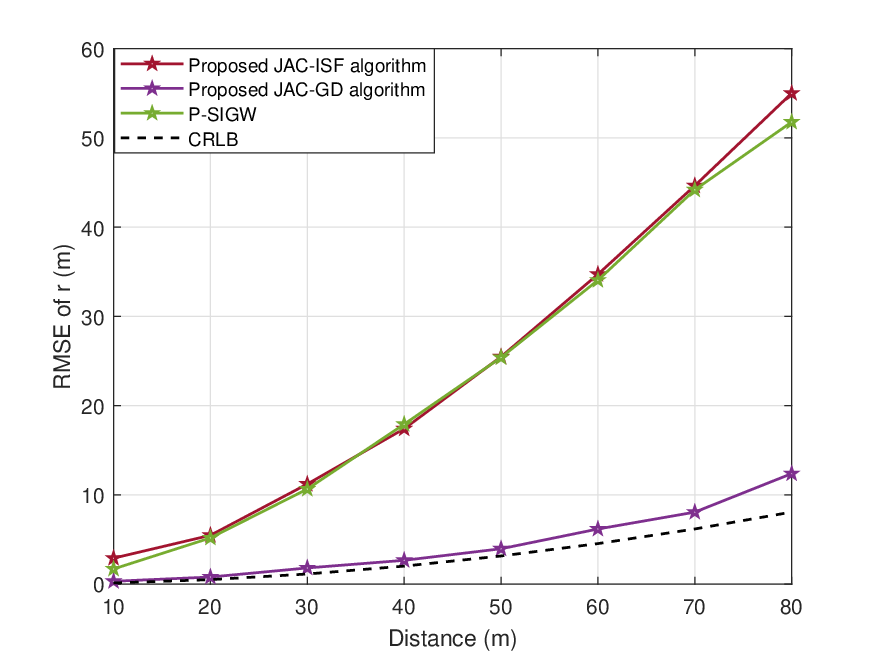}}
	\caption{Comparison of RMSE at different distances with ${\rm SNR}=5$.}
	\label{distance_RMSE}
\end{figure}

Fig. \ref{snr_RMSE}-\ref{distance_RMSE} show the RMSE performance of various algorithms at different SNR, snapshot points, and distance.  Due to the  CRLB depends on $r$ in equation (\ref{CRLBr}), we only present CRLB when comparing performance at different  positions. The Fig. \ref{snr_RMSE} shows the performance comparison at $T=8$. We can see that the JAC-GD algorithm performs well at low SNR regime.  The JAC-ISF algorithm performs slightly better than P-SIGW in the low SNR range, but as the SNR increases, the trend of accuracy improvement in the JAC-ISF algorithm is not as fast as the other two algorithms. This shows that the JAC-ISF algorithm has performance advantages when resources are scarce, while the JAC-GD algorithm has performance advantages in most scenarios.  The conclusion of Fig. \ref{plength_RMSE} is similar to the above conclusion, that the JAC-ISF algorithm has performance advantages at less snapshot points, while the JAC-GD algorithm generally has performance advantages.  In Fig. \ref{distance_RMSE}, we present the estimation results of $r$ and $\theta$ separately, and plot the CRLB as a benchmark. We can see that the JAC-GD algorithm is closest to the CRLB. This demonstrates that the JAC-GD algorithm not only has low time complexity, but also has significant performance advantages.

\section{Conclusion}

In this paper, we proposed a novel model and algorithm for near-field uplink channel estimation based on autocorrelation function, which solved the parameter coupling problem of near-field channel estimation with linear time complexity.  Specifically, we first decomposed the near-field channel and concluded that near-field channel estimation only requires estimating the parameters $p_1$ and $p_2$. By analyzing the physical meanings of $p_1$ and $p_2$, we found that essentially we need to estimate the CoA and AoA information of the incoming wave.  Through deduction, we found that the estimation problem of CoA is similar to that of Doppler. Therefore, by analogy with the definition of coherence time, we provided the definition of the coherence space of a channel. Based on the above analysis, we got the conclusion that $p_1$ can be solved by computing the autocorrelation function of the received signal. By solving the autocorrelation function, we found that it is only related to $p_1$ and independent of $p_2$, which means we successfully decoupled $p_1$ and $p_2$. After estimating $p_1$, the process of estimating $p_2$ degenerated into a far-field channel estimation problem. Thus, we used the classical MUSIC algorithm to estimate $p_2$. Based on the above model, we proposed two algorithms, i.e. the JAC-ISF and JAC-GD algorithm. Extensive simulation results under various practical setups demonstrated that the JAC-ISF and JAC-GD algorithm has lower time complexity compared to other algorithms and can achieved better performance with less snapshot points. In the future, we will design near-field algorithms for more complex scenarios based on the JAC model.

\ifCLASSOPTIONcaptionsoff
\newpage
\fi

\bibliographystyle{IEEEtran} 
\bibliography{myref}  

\begin{thebibliography}{10}
\providecommand{\url}[1]{#1}
\csname url@samestyle\endcsname
\providecommand{\newblock}{\relax}
\providecommand{\bibinfo}[2]{#2}
\providecommand{\BIBentrySTDinterwordspacing}{\spaceskip=0pt\relax}
\providecommand{\BIBentryALTinterwordstretchfactor}{4}
\providecommand{\BIBentryALTinterwordspacing}{\spaceskip=\fontdimen2\font plus
\BIBentryALTinterwordstretchfactor\fontdimen3\font minus \fontdimen4\font\relax}
\providecommand{\BIBforeignlanguage}[2]{{%
\expandafter\ifx\csname l@#1\endcsname\relax
\typeout{** WARNING: IEEEtran.bst: No hyphenation pattern has been}%
\typeout{** loaded for the language `#1'. Using the pattern for}%
\typeout{** the default language instead.}%
\else
\language=\csname l@#1\endcsname
\fi
#2}}
\providecommand{\BIBdecl}{\relax}
\BIBdecl

\bibitem{DBLP:journals/bell/Marzetta15}
T.~L. Marzetta, ``Massive {MIMO:} an introduction,'' \emph{Bell Labs Tech. J.}, vol.~20, pp. 11--22, Mar. 2015.

\bibitem{9184098}
E.~Björnson and L.~Sanguinetti, ``Power scaling laws and near-field behaviors of massive {MIMO} and intelligent reflecting surfaces,'' \emph{IEEE Open Journal of the Communications Society}, vol.~1, pp. 1306--1324, Sep. 2020.

\bibitem{DBLP:journals/dsp/BjornsonSWHM19}
E.~Bj{\"{o}}rnson, L.~Sanguinetti, H.~Wymeersch, J.~Hoydis, and T.~L. Marzetta, ``Massive {MIMO} is a reality - what is next?: Five promising research directions for antenna arrays,'' \emph{Digit. Signal Process.}, vol.~94, pp. 3--20, Nov. 2019.

\bibitem{8644126}
A.~Amiri, M.~Angjelichinoski, E.~de~Carvalho, and R.~W. Heath, ``Extremely large aperture massive {MIMO}: Low complexity receiver architectures,'' in \emph{2018 IEEE Globecom Workshops (GC Wkshps)}, Dec. 2018.

\bibitem{8948302}
H.~Wang, A.~Kosasih, C.-K. Wen, S.~Jin, and W.~Hardjawana, ``Expectation propagation detector for extra-large scale massive {MIMO},'' \emph{IEEE Transactions on Wireless Communications}, vol.~19, no.~3, pp. 2036--2051, Mar. 2020.

\bibitem{DBLP:journals/icl/WeiD22}
X.~Wei and L.~Dai, ``Channel estimation for extremely large-scale massive {MIMO:} far-field, near-field, or hybrid-field?'' \emph{{IEEE} Commun. Lett.}, vol.~26, no.~1, pp. 177--181, Nov. 2022.

\bibitem{DBLP:journals/tcom/0013Z0LHY24}
X.~Zhang, H.~Zhang, J.~Zhang, C.~Li, Y.~Huang, and L.~Yang, ``Codebook design for extremely large-scale {MIMO} systems: Near-field and far-field,'' \emph{{IEEE} Trans. Commun.}, vol.~72, no.~2, pp. 1191--1206, Feb. 2024.

\bibitem{DBLP:journals/comsur/WangYGZLZWHCHTLRTZSPH23}
C.~Wang, X.~You, X.~Gao, X.~Zhu, Z.~Li, C.~Zhang, H.~Wang, Y.~Huang, Y.~Chen, H.~Haas, J.~S. Thompson, E.~G. Larsson, M.~D. Renzo, W.~Tong, P.~Zhu, X.~Shen, H.~V. Poor, and L.~Hanzo, ``On the road to {6G}: Visions, requirements, key technologies, and testbeds,'' \emph{{IEEE} Commun. Surv. Tutorials}, vol.~25, no.~2, pp. 905--974, Feb. 2023.

\bibitem{9324910}
N.~Shlezinger, G.~C. Alexandropoulos, M.~F. Imani, Y.~C. Eldar, and D.~R. Smith, ``Dynamic metasurface antennas for {6G} extreme massive {MIMO} communications,'' \emph{IEEE Wireless Communications}, vol.~28, no.~2, pp. 106--113, Jan. 2021.

\bibitem{9933498}
N.~Yang and A.~Shafie, ``Terahertz communications for massive connectivity and security in {6G} and beyond era,'' \emph{IEEE Communications Magazine}, vol.~62, no.~2, pp. 72--78, Feb. 2022.

\bibitem{9665432}
Z.~Chen, C.~Han, Y.~Wu, L.~Li, C.~Huang, Z.~Zhang, G.~Wang, and W.~Tong, ``Terahertz wireless communications for 2030 and beyond: A cutting-edge frontier,'' \emph{IEEE Communications Magazine}, vol.~59, no.~11, pp. 66--72, Nov. 2021.

\bibitem{10149471}
Y.~Pan, C.~Pan, S.~Jin, and J.~Wang, ``{RIS}-aided near-field localization and channel estimation for the {Terahertz} system,'' \emph{IEEE Journal of Selected Topics in Signal Processing}, vol.~17, no.~4, pp. 878--892, Jun. 2023.

\bibitem{DBLP:journals/twc/ZhengZ23}
B.~Zheng and R.~Zhang, ``Simultaneous transmit diversity and passive beamforming with large-scale intelligent reflecting surface,'' \emph{{IEEE} Trans. Wirel. Commun.}, vol.~22, no.~2, pp. 920--933, Feb. 2023.

\bibitem{DBLP:journals/twc/WuZ19}
Q.~Wu and R.~Zhang, ``Intelligent reflecting surface enhanced wireless network via joint active and passive beamforming,'' \emph{{IEEE} Trans. Wirel. Commun.}, vol.~18, no.~11, pp. 5394--5409, Nov. 2019.

\bibitem{DBLP:conf/iccchina/PengZJLL23}
X.~Peng, L.~Zhao, Y.~Jiang, J.~Liu, and W.~Li, ``Channel estimation for extremely large-scale massive {MIMO} systems in hybrid-field channel,'' in \emph{{IEEE/CIC} International Conference on Communications in China, {ICCC} 2023, Dalian, China, August 10-12, 2023}.\hskip 1em plus 0.5em minus 0.4em\relax {IEEE}, Aug. 2023, pp. 1--6.

\bibitem{DBLP:conf/acssc/TorresSB20}
A.~de~Jesus~Torres, L.~Sanguinetti, and E.~Bj{\"{o}}rnson, ``Near- and far-field communications with large intelligent surfaces,'' in \emph{54th Asilomar Conference on Signals, Systems, and Computers, {ACSCC} 2020, Pacific Grove, CA, USA, November 1-4, 2020}, M.~B. Matthews, Ed.\hskip 1em plus 0.5em minus 0.4em\relax {IEEE}, Nov. 2020, pp. 564--568.

\bibitem{9743350}
H.~Zhang, N.~Shlezinger, F.~Guidi, D.~Dardari, M.~F. Imani, and Y.~C. Eldar, ``Near-field wireless power transfer for {6G} internet of everything mobile networks: Opportunities and challenges,'' \emph{IEEE Communications Magazine}, vol.~60, no.~3, pp. 12--18, Mar. 2022.

\bibitem{9903389}
M.~Cui, Z.~Wu, Y.~Lu, X.~Wei, and L.~Dai, ``Near-field {MIMO} communications for {6G}: Fundamentals, challenges, potentials, and future directions,'' \emph{IEEE Communications Magazine}, vol.~61, no.~1, pp. 40--46, Sep. 2023.

\bibitem{7942128}
K.~T. Selvan and R.~Janaswamy, ``Fraunhofer and fresnel distances: Unified derivation for aperture antennas,'' \emph{IEEE Antennas and Propagation Magazine}, vol.~59, no.~4, pp. 12--15, Aug. 2017.

\bibitem{DBLP:conf/wcsp/WangZZTY21}
T.~Wang, K.~Zhang, Y.~Zhang, H.~Tong, and C.~Yin, ``Near-field beam management in lis-assisted {mmWave} systems,'' in \emph{13th International Conference on Wireless Communications and Signal Processing, {WCSP} 2021, Changsha, China, October 20-22, 2021}.\hskip 1em plus 0.5em minus 0.4em\relax {IEEE}, Oct. 2021, pp. 1--6.

\bibitem{DBLP:conf/pimrc/HuIW22}
S.~Hu, M.~C. Ilter, and H.~Wang, ``Near-field beamforming for large intelligent surfaces,'' in \emph{2022 {IEEE} 33rd Annual International Symposium on Personal, Indoor and Mobile Radio Communications (PIMRC), Kyoto, Japan, September 12-15, 2022}.\hskip 1em plus 0.5em minus 0.4em\relax {IEEE}, Jun. 2022, pp. 1367--1373.

\bibitem{2023Extremely}
F.~Zheng, ``Extremely large-scale array systems: Near-filed codebook design and performance analysis,'' \emph{ArXiv}, vol. abs/2306.01458, 2023.

\bibitem{10005200}
Y.~Xie, B.~Ning, L.~Li, and Z.~Chen, ``Near-field beam training in {THz} communications: The merits of uniform circular array,'' \emph{IEEE Wireless Communications Letters}, vol.~12, no.~4, pp. 575--579, Jan. 2023.

\bibitem{10146329}
S.~Hu, H.~Wang, and M.~C. Ilter, ``Design of near-field beamforming for large intelligent surfaces,'' \emph{IEEE Transactions on Wireless Communications}, vol.~23, no.~1, pp. 762--774, Jun. 2023.

\bibitem{DBLP:journals/tsp/Friedlander19a}
B.~Friedlander, ``Localization of signals in the near-field of an antenna array,'' \emph{{IEEE} Trans. Signal Process.}, vol.~67, no.~15, pp. 3885--3893, Aug. 2019.

\bibitem{10476971}
A.~Kosasih, Ã.~T. Demir, and E.~Björnson, ``Parametric near-field channel estimation for extremely large aperture arrays,'' in \emph{2023 57th Asilomar Conference on Signals, Systems, and Computers}, Oct. 2023, pp. 162--166.

\bibitem{10132547}
X.~Guo, Y.~Chen, and Y.~Wang, ``Compressed channel estimation for near-field {XL-MIMO} using triple parametric decomposition,'' \emph{IEEE Transactions on Vehicular Technology}, vol.~72, no.~11, pp. 15\,040--15\,045, May. 2023.

\bibitem{DBLP:conf/acssc/BjornsonDS21}
E.~Bj{\"{o}}rnson, {\"{O}}.~T. Demir, and L.~Sanguinetti, ``A primer on near-field beamforming for arrays and reconfigurable intelligent surfaces,'' in \emph{55th Asilomar Conference on Signals, Systems, and Computers, {ACSSC} 2021, Pacific Grove, CA, USA, October 31 - November 3, 2021}.\hskip 1em plus 0.5em minus 0.4em\relax {IEEE}, Oct. 2021, pp. 105--112.

\bibitem{DBLP:journals/twc/JiangGJZZ23}
Y.~Jiang, F.~Gao, M.~Jian, S.~Zhang, and W.~Zhang, ``Reconfigurable intelligent surface for near field communications: Beamforming and sensing,'' \emph{{IEEE} Trans. Wirel. Commun.}, vol.~22, no.~5, pp. 3447--3459, May. 2023.

\bibitem{DBLP:journals/tsp/Collier05}
S.~L. Collier, ``Fisher information for a complex gaussian random variable: beamforming applications for wave propagation in a random medium,'' \emph{{IEEE} Trans. Signal Process.}, vol.~53, no.~11, pp. 4236--4248, November. 2005.

\bibitem{1994Fundamentals}
S.~K. Sengijpta, ``Fundamentals of statistical signal processing: Estimation theory,'' \emph{Control Engineering Practice}, vol.~37, no.~4, pp. 465--466, 1994.

\bibitem{DBLP:journals/tcom/CuiD22}
M.~Cui and L.~Dai, ``Channel estimation for extremely large-scale {MIMO:} far-field or near-field?'' \emph{{IEEE} Trans. Commun.}, vol.~70, no.~4, pp. 2663--2677, Apr. 2022.

\bibitem{9802728}
L.~Li, H.~Li, Z.~Chen, W.~Chen, and S.~Li, ``An analytical range-angle dependent beam focusing model for {Terahertz} linear antenna array,'' \emph{IEEE Wireless Communications Letters}, vol.~11, no.~9, pp. 1870--1874, Jun. 2022.

\bibitem{8306126}
J.~Rodríguez-Fernández, N.~González-Prelcic, K.~Venugopal, and R.~W. Heath, ``Frequency-domain compressive channel estimation for frequency-selective hybrid millimeter wave {MIMO} systems,'' \emph{IEEE Transactions on Wireless Communications}, vol.~17, no.~5, pp. 2946--2960, Mar. 2018.

\bibitem{9246294}
N.~González-Prelcic, H.~Xie, J.~Palacios, and T.~Shimizu, ``Wideband channel tracking and hybrid precoding for {mmWave MIMO} systems,'' \emph{IEEE Transactions on Wireless Communications}, vol.~20, no.~4, pp. 2161--2174, Oct. 2021.

\end{thebibliography}
\end{document}